\documentclass[10pt,twocolumn,showpacs,amsmath,amssymb,floatfix,superscriptaddress]{revtex4-2}

\usepackage[T1]{fontenc}
\usepackage{graphicx}
\usepackage{dcolumn}
\usepackage{bm}
\usepackage{color}
\usepackage{txfonts}
\usepackage{microtype}
\usepackage{hyperref}
\usepackage[english]{babel}
\usepackage{slashed}
\usepackage{gensymb}
\usepackage{epsfig}
\usepackage[normalem]{ulem}
\usepackage{amsmath}
\usepackage{array}
\usepackage{booktabs}
\usepackage{nccmath}
\allowdisplaybreaks[4]

\begin{document}
\renewcommand{\figurename}{FIG}	
	
\title{Isolated Attosecond $\gamma$-Ray Pulse Generation with Transverse Orbital Angular Momentum Using Intense Spatiotemporal Optical Vortex Lasers}

\author{Fengyu Sun}\thanks{These authors have contributed equally to this work.}
\affiliation{State Key Laboratory of High Field Laser Physics and CAS Center for Excellence in Ultra-intense Laser Science, Shanghai Institute of Optics and Fine Mechanics (SIOM), Chinese Academy of Sciences (CAS), Shanghai 201800, China}
\affiliation{School of Physical Science and Technology, ShanghaiTech University, Shanghai 201210, China}
\affiliation{University of Chinese Academy of Sciences, Beijing 100049, China}	

\author{Xinyu Xie}\thanks{These authors have contributed equally to this work.}
\affiliation{State Key Laboratory of High Field Laser Physics and CAS Center for Excellence in Ultra-intense Laser Science, Shanghai Institute of Optics and Fine Mechanics (SIOM), Chinese Academy of Sciences (CAS), Shanghai 201800, China}
\affiliation{University of Chinese Academy of Sciences, Beijing 100049, China}	

\author{Wenpeng Wang}\email{wangwenpeng@siom.ac.cn}
\affiliation{State Key Laboratory of High Field Laser Physics and CAS Center for Excellence in Ultra-intense Laser Science, Shanghai Institute of Optics and Fine Mechanics (SIOM), Chinese Academy of Sciences (CAS), Shanghai 201800, China}

\author{Stefan Weber}
\affiliation{ELI Beamlines facility, Extreme Light Infrastructure ERIC, 25241 Dolni Brezany, Czech Republic}

\author{Xin Zhang}  	
\affiliation{State Key Laboratory of High Field Laser Physics and CAS Center for Excellence in Ultra-intense Laser Science, Shanghai Institute of Optics and Fine Mechanics (SIOM), Chinese Academy of Sciences (CAS), Shanghai 201800, China}

\author{Yuxin Leng}
\affiliation{State Key Laboratory of High Field Laser Physics and CAS Center for Excellence in Ultra-intense Laser Science, Shanghai Institute of Optics and Fine Mechanics (SIOM), Chinese Academy of Sciences (CAS), Shanghai 201800, China}

\author{Ruxin Li}\email{ruxinli@siom.ac.cn}
\affiliation{State Key Laboratory of High Field Laser Physics and CAS Center for Excellence in Ultra-intense Laser Science, Shanghai Institute of Optics and Fine Mechanics (SIOM), Chinese Academy of Sciences (CAS), Shanghai 201800, China}
\affiliation{School of Physical Science and Technology, ShanghaiTech University, Shanghai 201210, China}

\author{Zhizhan Xu}
\affiliation{State Key Laboratory of High Field Laser Physics and CAS Center for Excellence in Ultra-intense Laser Science, Shanghai Institute of Optics and Fine Mechanics (SIOM), Chinese Academy of Sciences (CAS), Shanghai 201800, China}

\date{\today}

\begin{abstract}
An isolated attosecond vortex $\gamma$-ray pulse is generated by using a relativistic spatiotemporal optical vortex (STOV) laser in particle-in-cell simulations. A $\sim$ 300-attosecond electron slice with transverse orbital angular momentum (TOAM) is initially selected and accelerated by the central spatiotemporal singularity of the STOV laser. This slice then collides with the laser's reflected Gaussian-like front from a planar target, initiating nonlinear Compton scattering and resulting in an isolated, attosecond ($\sim$ 300 as), highly collimated ($\sim$ 4$\degree$), ultra-brilliant ($\sim 5\times 10^{24}$ photons/s/mm$^2$/mrad$^2$/0.1\%BW at 1 MeV) $\gamma$-ray pulse. This STOV-driven approach overcomes the significant beam divergence and complex two-laser requirements of prior Gaussian-based methods while introducting TOAM to the attosecond $\gamma$-ray pulse, which opens avenues for ultrafast imaging, nuclear excitation, and detection applications.

\end{abstract}

\maketitle
The 2023 Noble Prize was awarded to Agostini, Krausz and L'Huillier for their pioneering work on generation of attosecond pulses \cite{RevModPhys.96.030501,RevModPhys.96.030502,RevModPhys.96.030503}. Attosecond pulses were initially generated through high harmonic generation (HHG) using nonrelativistic laser-driven gas targets \cite{Sansone2011,Yeung2017,RevModPhys.81.445}. However, this approach typically results in a pulse train due to the multi-cycle nature of the driving laser \cite{PhysRevLett.77.1234,doi:10.1126/science.1059413}, rather than an isolated pulse, which is crucial for probing attosecond phenomena in electron physics by enabling undisturbed, single-shot observations \cite{doi:10.1126/science.1080552,PhysRevLett.124.114802}. To address this, several methods have been proposed for generating isolated attosecond pulses, including the use of few-cycle laser pulses \cite{doi:10.1126/science.1059413}, polarization gating \cite{PhysRevLett.112.123902,Rykovanov_2008}, and other gating techniques \cite{Rykovanov_2008,PhysRevLett.108.235003}. Despite these advances, the generated attosecond pulses were initially limited to the mid-infrared range, only extending into the soft X-ray region \cite{RevModPhys.81.163}.

For higher-energy pulses, $\gamma$-ray pulses with $\sim$MeV-level photons have garnered increasing interest due to their potential applications in nuclear physics \cite{doi:10.1126/science.1080552,PhysRevLett.115.204801}, as typical reaction energies within the nucleus occur at the MeV scale \cite{Nedorezov_2021}. Traditionally, $\gamma$-rays have been efficiently generated via nonlinear Thomson/Compton scattering \cite{Gu2018,PhysRevLett.113.224801,Cipiccia2011,TaPhuoc2012} and nonlinear synchrotron radiation \cite{doi:10.1073/pnas.1809649115,doi:10.1126/sciadv.aaz7240} driven by relativistic lasers. To produce attosecond $\gamma$-rays, an approach using Gaussian lasers was initially proposed, where an isolated attosecond electron slice (IAES) is directly accelerated from an ultra-thin target and subsequently collides with an additional relativistic laser to generate an isolated attosecond $\gamma$-ray pulse \cite{Zhang_2022}. However, this scheme faces two major challenges: the first is the significant divergence and low brightness of the $\gamma$-ray pulse, caused by the transverse ponderomotive force inherent to Gaussian laser fields \cite{PhysRevLett.109.115002,Thévenet2016}; the second is the reliance on a two-laser system, which requires precise spatiotemporal synchronization between the lasers \cite{Li:17,Zhao2022}. To address the first issue, structured laser pulses, such as Laguerre-Gaussian (LG) lasers, have been proposed. These lasers, with their hollow intensity distribution, can generate attosecond $\gamma$-ray trains with reduced divergence and enhanced brightness \cite{PhysRevA.45.8185,PhysRevLett.122.024801,Wang2015,PhysRevApplied.12.014054}. Unfortunately, these pulses are not isolated. A natural progression is to add attosecond temporal confinement to the spatial-vortex LG lasers, which could potentially resolve this limitation.

Fortunately, the spatiotemporal optical vortex (STOV) pulse possesses the desired characteristics, extending the vortex structure from the spatial domain into the spatiotemporal realm \cite{PhysRevA.86.033824,PhysRevX.6.031037,Chong2020,Wan2022}. This pulse offers three promising advantages for attosecond $\gamma$-ray generation. First, the extremely small spatiotemporal singularity at the center of the STOV laser acts as an "attosecond selector" \cite{PhysRevResearch.6.013075}, enabling ultra-short, collimated particle modulation. Second, the front of the STOV laser exhibits a near-Gaussian intensity profile, which can be reflected and made to collide with particles at the singularity within a single-laser system, potentially simplifying experimental setups. Third, unlike traditional Gaussian or LG lasers, the STOV pulse imparts a novel transverse orbital angular momentum (TOAM) to the attosecond $\gamma$-ray, offering new dimensions for detection and triggering tools in various applications.

In this letter, we propose using an intense circularly polarized (CP) STOV laser to generate a collimated, ultrabright, and isolated attosecond $\gamma$-ray pulse with TOAM in three-dimensional (3D) particle-in-cell (PIC) simulations. Electrons are extracted and accelerated by the STOV laser's fields, forming an IAES. This slice collides with the reflected Gaussian-like laser front from a planar target to trigger the nonlinear Compton scattering (NCS) process, producing a highly collimated ($\sim$4$\degree$) ultra-brilliant (~$\sim5\times10^{24}$ photons/s/mm$^2$/mrad$^2$/0.1\%BW at 1 MeV), and attosecond ($\sim$300 as) $\gamma$-ray pulse using $\sim10^{22}$ W/cm$^2$ laser intensity. Notably, TOAM is successfully transferred from the STOV laser to the $\gamma$-ray pulse, providing new opportunities for ultrahigh time resolution observation \cite{Drescher2002}, nuclear selective excitation \cite{PhysRevLett.131.202502}, and detection in nuclear science \cite{doi:10.1126/science.1080552,Habs2009}. 

\begin{figure}[h]
	\setlength{\abovecaptionskip}{0.25cm}
	\centering\includegraphics[width=1\linewidth]{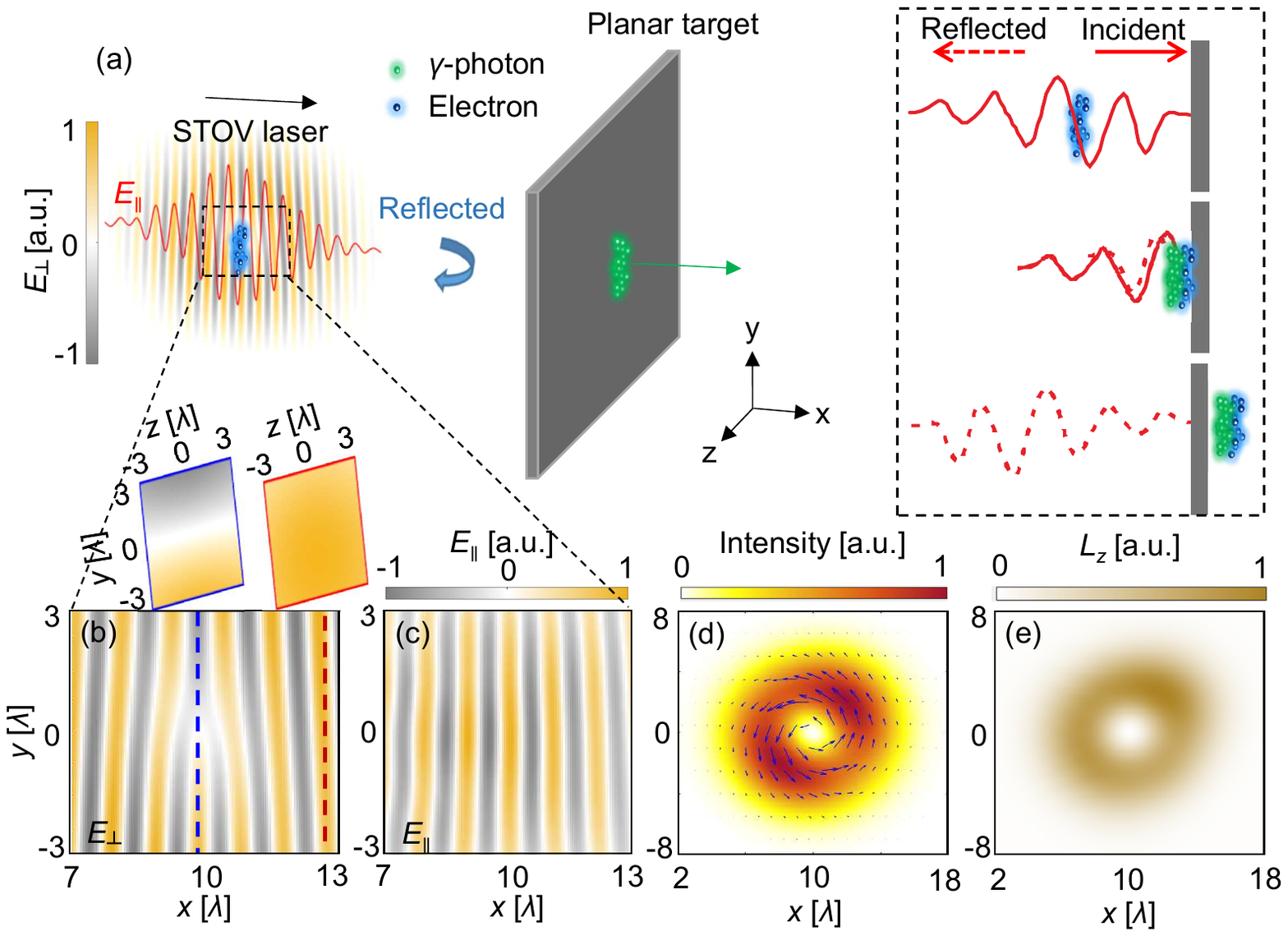}
	\caption{(a) Schematic of an isolated attosecond $\gamma$-ray pulse generation. An IAES collides with the self-reflected STOV laser, triggering the NCS process to generate the attosecond $\gamma$-ray pulse. (b) Vertical electric field and (c) longitudinal electric field distributions of the STOV laser. (d) The energy density of the STOV laser, with blue arrows indicating the circulating momentum flux. (e) TOAM density with the propagation term subtracted.}
	\label{Fig1}
\end{figure}
The generation of attosecond $\gamma$-ray pulse driven by the STOV laser was simulated using the 3D PIC code EPOCH \cite{Arber_2015,RIDGERS2014273} [see Fig. \ref{Fig1} (a)]. An intense STOV laser, with a duration of $\sim$ 30 fs, propagates along the $+x$ axis from the left side of the simulation box. The vertical electric field of the STOV laser pulse can be expressed as [see Fig. \ref{Fig1} (b)]:
\begin{equation}
	\begin{split}
		\mathbf{E}_{\bot}&=E_0\frac{w_0}{w}\left( \frac{\xi ^2+y^2}{w^2} \right) ^{\frac{\left| l \right|}{2}}\exp \left( -\frac{\xi ^2+y^2+z^2}{w^2} \right) \times 
		\\
		&\exp \left[ i\left( -l\phi _{\text{st}}+kx-\omega t+\phi _0 \right) \right] \exp \left( \mathbf{\hat{e}}_y+i\pi\sigma _z/2\cdot\mathbf{\hat{e}}_z \right) \label{eq1},
	\end{split}
\end{equation}
where $E_0=a_0m_\text{e}c\omega/e$ is the peak amplitude of the electric field, $a_0=70$ is the normalized laser amplitude (corresponding to the laser intensity $I_0=1.35\times10^{22}$ W/cm$^2$, which can be achieved under the current laboratory conditions \cite{Li:18,Danson_Haefner_Bromage_Butcher_Chanteloup_Chowdhury_Galvanauskas_Gizzi_Hein_Hillier_etal._2019,Yoon:21}). The longitudinal coordinate local to the laser is $\xi=ct-x$, where $c$ is the speed of light in vacuum. $\omega=4\lambda$ is the beam waist, $\lambda=1$ $\mu$m is the laser center wavelength. $l=1$ is the topological charge number, $\phi_{\text{st}}= \tan^{-1}(y/\xi)$ is the spatiotemporal azimuthal angle, $k=2\pi/\lambda$ is the wave number, $\phi_0=0$ is the initial phase and $\sigma_z=-1$ is the spin quantum number. Spatial and temporal profiles are used with $\sqrt{\xi^2+y^2}$. The longitudinal electric field $E_{\parallel}$ can be calculated using the Poisson equation: $\mathbf{E}_{\parallel}=-(i/k)( \partial \mathbf{E}_{\bot}/\partial y+\partial \mathbf{E}_{\bot}/\partial z) $ [see Fig. \ref{Fig1} (c)].
\begin{equation}
	\begin{split}
		&\mathbf{E}_{\varparallel}=E_0\frac{i}{k}\frac{w_0}{w}\exp \left[ -\frac{\xi ^2+x^2+y^2}{w^2}+i\left( -l\phi _{\text{st}}+kx-\omega t+\phi _0 \right) \right]\times  
		\\
		&\exp \left( \mathbf{\hat{e}}_x \right)\left[ \left( \frac{\xi ^2+y^2}{w^2} \right) ^{\frac{l}{2}}\left( \frac{2z+2y}{w^2}+\frac{ix}{x^2+y^2} \right) -\frac{1}{2}\left( \frac{\xi ^2+y^2}{w^2} \right) ^{-\frac{l}{2}}\frac{2y}{w^2} \right].
	\end{split}
\end{equation}

Figure \ref{Fig1} (d) shows the time-averaged energy density of the STOV laser, which can be expresses as $I=( \varepsilon _0\left| \mathbf{E} \right|^2+\left| \mathbf{B} \right|^2/\mu _0) /4$ \cite{Zhang_Ji_Shen_2022,10.1063/5.0214297,Ju_2024}, where $\mathbf{E}$ and $\mathbf{B}$ represent the electric and magnetic fields of the STOV laser, while $\varepsilon _0$ and $\mu _0$ are the dielectric constant and permeability of vacuum, respectively. Additionally, by subtracting the propagating momentum $I/c$ in the $+x$ direction, the energy flux circulates counterclockwise around the energy singularity, which can be calculated by the momentum vector $\mathbf{P}_{\text{c}}=(P_x-I/c,P_y,P_z) $, where the canonical momentum density $\mathbf{P}=\text{Im}\left[ \varepsilon _0\mathbf{E}^*\cdot \left( \nabla \right) \mathbf{E}+\mathbf{B}^*\cdot \left( \nabla \right) \mathbf{B/}\mu _0 \right] /(4\omega) $ \cite{PhysRevLett.113.033601,Bliokh2014,Bliokh_2013}. With the circulation of canonical momentum, the TOAM can be calculated as $\mathbf{L}=\mathbf{r}\times \mathbf{P}_{\text{c}}$, where $\mathbf{r}$ is the local position relative to the energy centroid. By integrating the angular momentum density over the full space, the TOAM per photon for the STOV pulse is $\sim 0.99\hbar$ [see Fig. \ref{Fig1} (e)]

The wire target is positioned between 0 $\mu$m and 4 $\mu$m with a diameter of 400 nm, and the planar target is positioned between 40 $\mu$m and 41 $\mu$m with a transverse area of 16$\times$16 $\mu$m$^2$. Their initial density is 100 $n_c$, composed of fully ionized hydrogen ions and electrons, where $n_c = m_\text{e}\omega^2/4\pi e^2$ is the critical density, $m_\text{e}$ is the electron mass, $\omega$ is the laser frequency and $e$ is the electron charge. The dimensions of the window are 15 $\mu$m ($x$) $\times$ 16 $\mu$m ($y$) $\times$ 16 $\mu$m ($z$) with 600$\times$640$\times$640 cells. Each cell contains 50 macro-particles for the wire target and 8 macro-particles for the planar target. Absorption boundary conditions are employed for both particles and fields. 

\begin{figure}[h]
	\setlength{\abovecaptionskip}{0.3cm}
	\centering\includegraphics[width=1\linewidth]{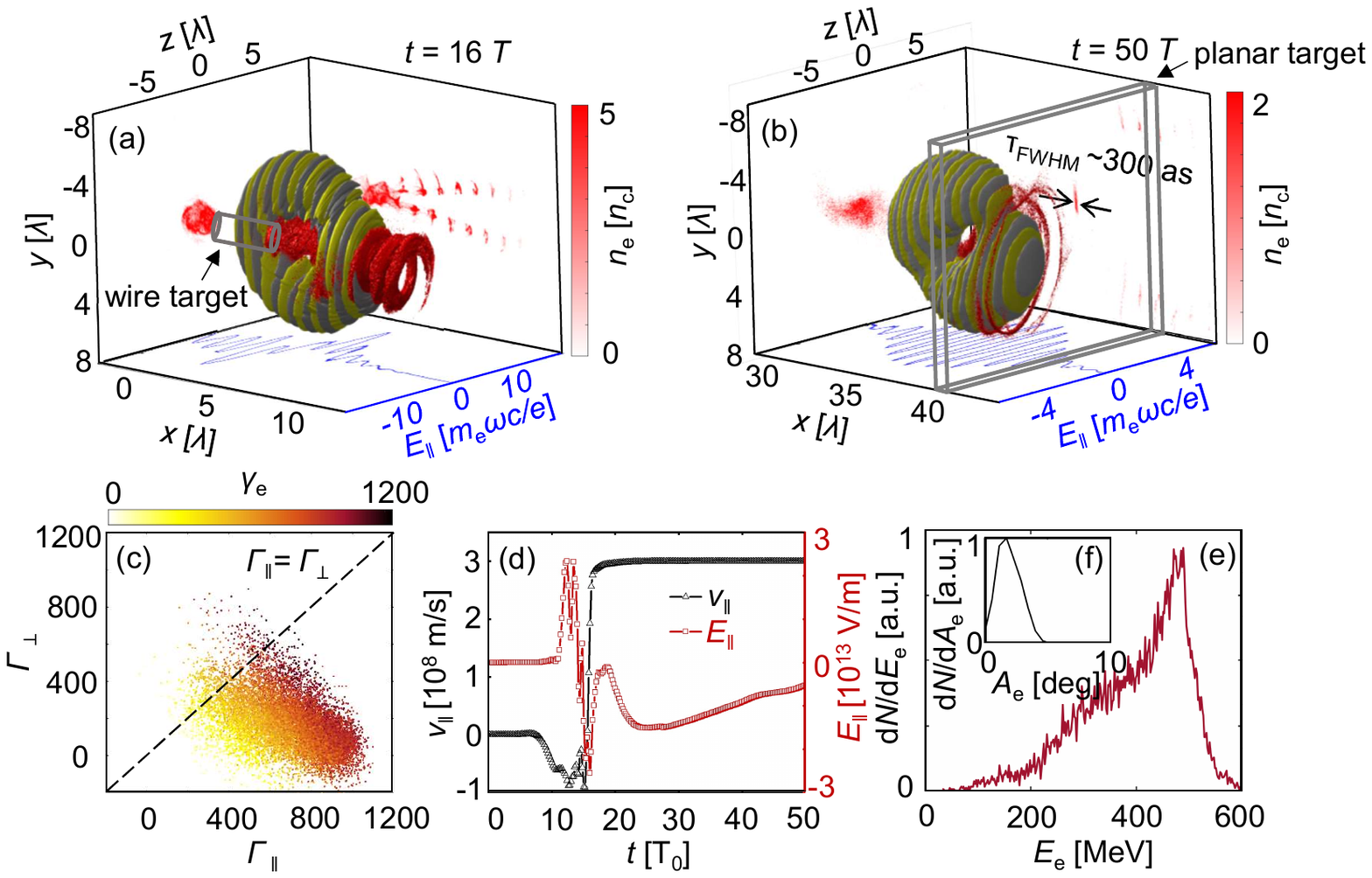}
	\caption{Energy density iso-surface at (a) $t =16 T_0$ ($T_0=\lambda/c$) with 500 MeV$\cdot$$n_c$ and (b) $t = 50 T_0$ with 1 GeV$\cdot$$n_c$; as well as the one-dimensional longitudinal electric field in the $x-y$ plane. (c) Energy-gain plane of $\Gamma _{\parallel}$ and $\Gamma _{\bot}$ of electrons, with their net $\gamma_\text{e}$-factor color-coded. (d) Average longitudinal electron speed $v_\parallel$ and experienced average longitudinal electric field $E_\parallel$ of the IAES as a function of $T_0$. (e) Energy spectrum and (f) divergence angle of the IAES.}
	\label{Fig2}
\end{figure}

When an intense STOV laser irradiates a wire target, electrons are expelled from the optical axis due to the transverse ponderomotive force $F_{\text{p}}=-m_{\text{e}}c^2/( 4\gamma _{\text{e}}) \nabla E^2(x,y)$. This occurs because the front of the vertical electric field of the STOV laser resembles that of a Gaussian mode laser [see Fig. \ref{Fig1} (b) and Fig. \ref{Fig2} (a)], where $E$ is the normalized laser amplitude derived from Eq. \ref{eq1}, $\gamma _{\text{e}}=(1-v^2/c^2)^{-1/2}$ is the relativistic parameter, $v$ is the electron speed. Consequently, only those electrons located at the laser phase singularity can be trapped within the laser field [see Fig. \ref{Fig2} (b)]. Additionally, the longitudinal electric field ($E_{\parallel}$) establishes an accelerating configuration within the hollow structure, enabling the acceleration of trapped electrons.

To investigate the acceleration mechanism in detail, Fig. \ref{Fig2} (c) presents the energy gain plane ($\Gamma _{\parallel}, \Gamma _{\bot}$) for the IAES, where $\Gamma _{\parallel}=-\int_0^t{ev_{\parallel}E_{\parallel}}\text{d}t$ and $\Gamma _{\bot}=-\int_0^t{ev_{\bot}E_{\bot}}\text{d}t$ denote the accumulated energy gains from the longitudinal and transverse fields, respectively \cite{Shen2024}. It is observed that the most energetic electrons are primarily accelerated by the longitudinal field, with the number of electrons accelerated longitudinally significantly exceeding that of those accelerated transversely. Therefore, the IAES is predominantly accelerated by the longitudinal electric field $E_{\parallel}$. It is noted that a brief yet intense longitudinal electric field of $E_\parallel\sim10^{13}$ V/m boosts the electron energy of $\gamma_\text{e}\sim30$, primarily driven by the strong longitudinal component $E_\parallel$ of the diffracted laser field at 16 $T_0$ [see Fig. \ref{Fig2} (d)]. After $t = 16 T_0$, such electrons traveling at speeds close to the light speed continue to accelerate to form the IAES within the same phase since the longitudinal field remains negative.

Finally, this phase-locked acceleration enables the IAES within the spatiotemporal phase singularity of the STOV laser to achieve a cut-off energy of $\sim$ 600 MeV [see Fig. \ref{Fig2} (e)], a charge of around 0.2 nC, and a divergence of about 2$\degree$ [see Fig. \ref{Fig2} (f)] with a full width at half maximum (FWHM) of $\sim$ 300 as. Unlike traditional Gaussian lasers, which cause the wire target to expel electrons due to the transverse pondermotive force, this force induces a divergence of the electrons without exhibiting localized properties.

\begin{figure}[h]
	\setlength{\abovecaptionskip}{0.3cm}
	\centering\includegraphics[width=1\linewidth]{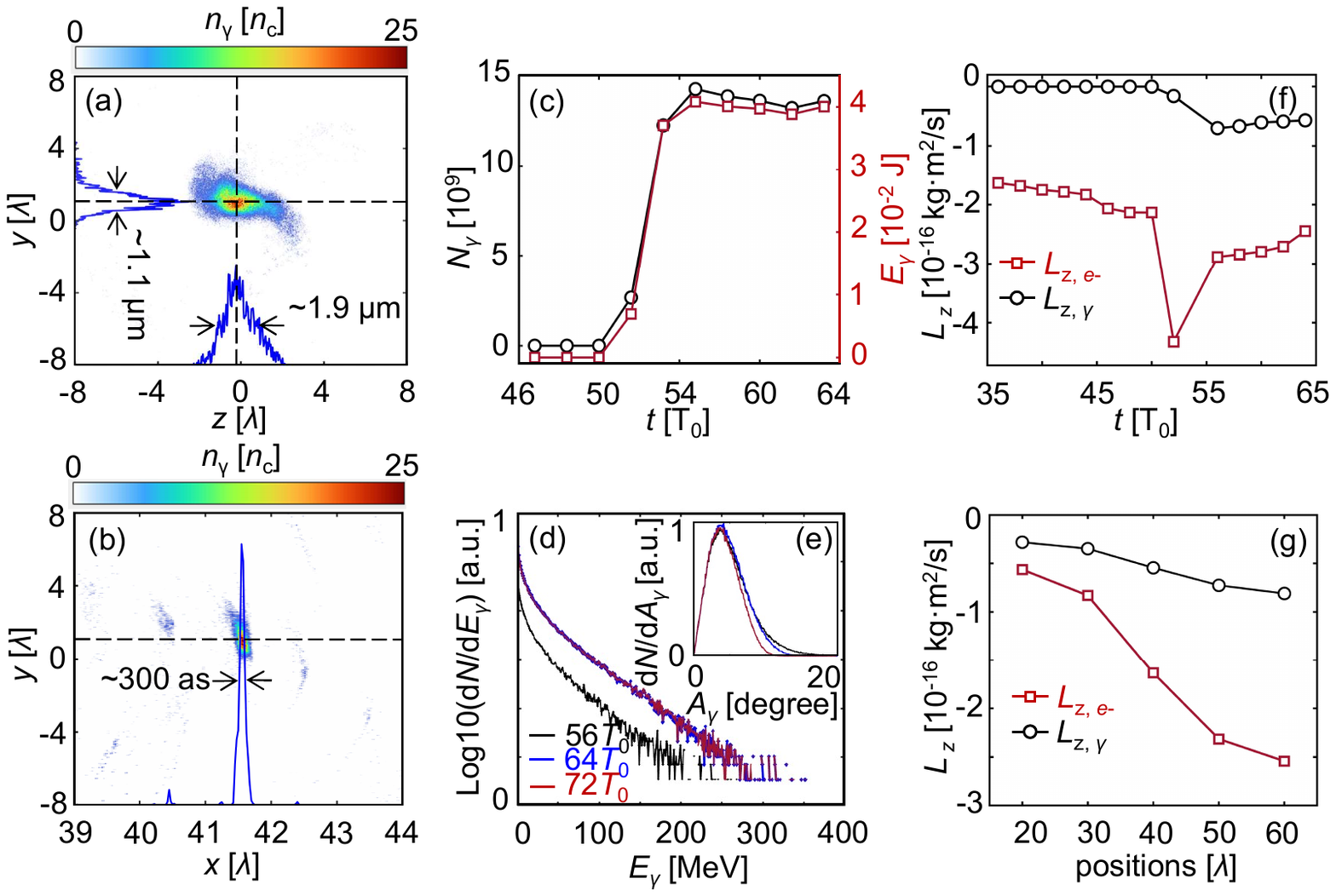}
	\caption{Density distribution of the isolated attosecond $\gamma$-ray pulse in the (a) $z-y$ plane and (b) $x-y$ plane at $t=56T_0$, with the blue line indicating the one-dimensional distribution along the line of black dashes. (c) Evolution of the photon number and the photon energy of the $\gamma$-ray pulse. (d) Energy spectra and (e) angular divergence of the $\gamma$-ray pulse at $t = 56$ $T_0$, 64 $T_0$, and 72 $T_0$. (f) Evolution of TOAM for the IAES and $\gamma$-ray pulse with the plane target at $x=40$ $\mu$m. (g) TOAM of the IAES and the $\gamma$-ray pulse at different plane target positions.}
	\label{Fig3}
\end{figure}
The hundreds-of-MeV IAES depicted in Fig. \ref{Fig2} demonstrate its efficacy as a particle source for subsequent attosecond $\gamma$-ray generation. In this process, a planar target reflects the Gaussian-like front of the STOV laser, resulting in a collision with the IAES within the singularity of the STOV laser. This interaction triggers the NCS process, leading to the generation of an isolated attosecond $\gamma$-ray pulse [see Figs. \ref{Fig3} (a) and (b)] \cite{PhysRevLett.113.224801,TaPhuoc2012}.

To facilitate the efficient emission of attosecond $\gamma$-rays, the quantum electrodynamics (QED) effects in photon radiation are characterized by the nonlinear QED parameter $\chi _{\text{e}}\equiv |e| \hbar/( m_{\text{e}}^{3}c^4) \sqrt{-( F_{\mu \nu}p^{\nu})}$ \cite{osti_5972043}, where $\chi _{\text{e}}\ge 0.1$ \cite{10.1063/1.5028555}. Here, $\hbar$ is the reduced Planck constant, $F_{\mu \nu}$ is the field tensor, and $p^{\nu}$ is the electron four-momentum. In our scheme, $\chi_{\text{e}}$ can be expressed by $\chi _{\text{e}}=\gamma _{\text{e}}\left| \mathbf{E}_{\bot}+\boldsymbol{\beta }\times c\mathbf{B} \right|/E_{\text{S}}$, where $\boldsymbol{\beta }=\boldsymbol{v}/c$ is the normalization electron velocity, $\mathbf{B}$ is the magnetic field, and $E_{\text{S}} = 1.3\times10^{18}$ V/m is the Schwinger field \cite{PhysRev.82.664}. 

When the laser propagates in the same direction as the IAES (along the $+x$ direction), $\mathbf{E}_{\bot}$ can be almost completely canceled by $\boldsymbol{\beta }\times c\mathbf{B}$, resulting in $\chi _{\text{e}}\simeq 0$. This condition inhibits the emission of high-energy $\gamma$ photons before $t = 50 T_0$. By contrast, after $t = 50 T_0$, the quantum parameter $\chi_{\text{e}}$ reaches 0.3 due to the collision between the reflected STOV laser (propagating in the $-x$ direction) and the high-energy IAES (propagating in the $+x$ direction), indicating a significant enhancement in attosecond $\gamma$-ray emission.

The power of the attosecond $\gamma$-ray emitted by the electron can be expressed as $P_{\text{rad}}\approx 2eE_{\text{S}}N_{\text{e}}c\alpha _f\chi _{\text{e}}^{2}g(\chi _{\text{e}}) /3$, where $N_{\text{e}}$ is the number of IAES, $\alpha _f$ is the fine-structure constant, and  $g\left( \chi _{\text{e}} \right) \approx ( 3.7\chi _{\text{e}}^{3}+31\chi _{\text{e}}^{2}+12\chi _{\text{e}}+1) ^{-4/9}$ is the radiation correction induced by the QED effects \cite{Duff_2018}. Starting from $t = 50 T_0$, as the instantaneous radiation power $P_{\text{rad}}$ increases, both the photon number and energy of the attosecond $\gamma$-ray experience significant growth [see Fig. \ref{Fig3} (c)]. By $t = 56 T_0$, the generation of $\gamma$ photons is essentially complete, marking the end of the NCS process. At this point, the photon energy and divergence angle of the attosecond $\gamma$-ray reach $\sim$ 300 MeV and $\sim$  70 mrad, respectively [see Figs. \ref{Fig3} (d) and (e)].

According to the pulse width of the IAES $\tau_\text{e}$ and the reflected STOV laser $\tau_\text{L}$, the generated pulse width of $\gamma$-ray can be described as $\tau_\gamma = \tau_\text{e}+\tau_\text{L}/(4\gamma_\text{e}^2)\approx$300 as \cite{PhysRevSTAB.3.090702,10.1063/1.4826600}, which is similar to the PIC simulation results [see Fig. \ref{Fig3} (b)]. Such an isolated $\gamma$-ray pulse, with the photon number exceeding $\sim10^{10}$ at 1 MeV, has a transverse dimension of about $1.1\times1.9 $ $\mu$m$^2$. This configuration results in an isolated collimated $\gamma$-ray source with the peak brilliance of $\sim 5\times 10^{24}$ photons/s/mm$^2$/mrad$^2$/0.1\%BW, significantly exceeding the brilliance achieved in current attosecond $\gamma$-ray studies \cite{PhysRevLett.113.224801,Zhang2021,10.1063/5.0028203}.

Since the STOV laser pulse carries an intrinsic TOAM $\sim 0.99\hbar$ value [see Fig. \ref{Fig1} (e)], there is the possibility for TOAM transfer during the generation of the isolated attosecond $\gamma$-ray pulse. The TOAM is an extrinsic property of particle pulses; thus, the TOAM of the IAES and the $\gamma$-ray pulse can be calculated in the laboratory frame using $L_z=\sum_i{(xp_y-yp_x)}$, where $i$ represents the index of the particles in the IAES and $\gamma$-ray pulse. The generated photons propagate in the direction of the electron motion, and their momentum originates from the parent electrons \cite{10.1063/1.5028555}. This interaction results in the generated isolated attosecond $\gamma$-ray carrying a TOAM of $\sim$ $0.5\times10^{-16}$ kg$\cdot$m$^2$/s, with the TOAM of the $\gamma$-ray pulse gradually stabilizing at the end of the interaction [see Fig. \ref{Fig3} (f)]. Such an isolated attosecond $\gamma$-ray pulse with TOAM presents new opportunities in nuclear physics and has garnered significant attention for its potential applications, such as modulating photonuclear reaction rates \cite{Afanasev_2018}, uncovering novel spin phenomena \cite{Afanasev_2017,PhysRevLett.124.192001}, disentangling spin states and dual resonances \cite{PhysRevD.101.016007}, and introducing innovative multipolar analyses in photonuclear reactions \cite{PhysRevA.97.023422}.

From the above analysis, it is clear that the TOAM of the attosecond $\gamma$-ray pulse originates from the TOAM of the colliding IAES, which, in turn, is imparted by the STOV laser. To further confirm this phenomenon, a planar target can be placed at a distance of 20 $\mu$m to 60 $\mu$m to reflect the STOV laser, thus influencing the interaction time between the STOV laser and the IAES [see Fig. \ref{Fig3} (g)]. During the electron acceleration process, the TOAM of the electrons increases as the the distance between the wire target and the planar target grows. This increase results from the continuous transfer of TOAM from the STOV laser to the IAES. Morever, in the NCS process, the electrons transfer their TOAM to the $\gamma$ photons, leading to a corresponding increase in the TOAM of the $\gamma$ photons as the electron TOAM rises. 

\begin{figure}[h]
	\setlength{\abovecaptionskip}{0.3cm}
	\centering\includegraphics[width=1\linewidth]{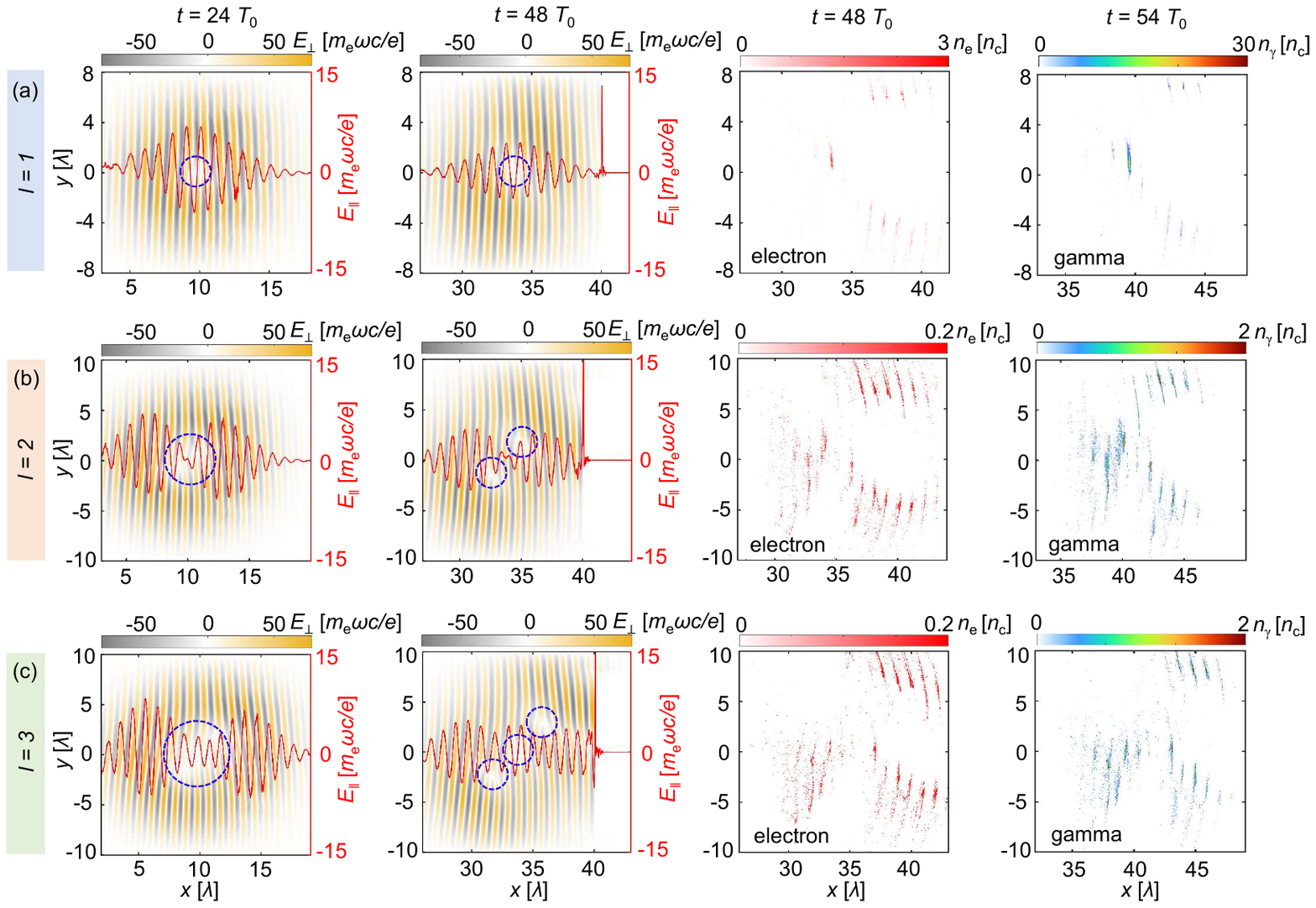}
	\caption{Vertical electric field $E_\bot$ and one-dimensional longitudinal electric field $E_\parallel$ on axis at $t = 24 T_0$ and $t = 48 T_0$; electron distribution in the $x-y$ plane at $t = 48 T_0$; $\gamma$-photon distribution in the $x-y$ plane at $t = 54 T_0$ driven by the (a) $l = 1$, (b) 2 and (c) 3 of the STOV laser. The blue dashed circle indicates the positions of the phase singularity.}
	\label{Fig4}
\end{figure}

It is important to note that the spatiotemporal phase singularity structure is also present at the laser center for STOV lasers with higher-order topological charges ($l=2$ and $l=3$) [see Figs. \ref{Fig4} (b) and (c)]. Unlike the case of $l=1$ [see Fig. \ref{Fig4} (a)], the phase difference between the upper and lower laser stripes is 4$\pi$ and 6$\pi$ for $l=2$ and $l=3$, respectively, generating $l$ electron slices within the phase singularity. Moreover, due to the spatial diffraction and structural instability, the higher-order STOV laser field evolves from one single singularity to $l$ singularities during laser propagation \cite{Porras:23}. This reduces the convergence effect of the vertical electric field on the electron slices, leading to increased transverse electron divergence. These electron slices produce multiple diverging attosecond $\gamma$-ray pulses through the NCS process. It is also noteworthy that for higher-order STOV fields ($l = 2$ and $l = 3$), the longitudinal electric field $E_\parallel$ within the phase singularity is weaker compared to the $l = 1$ case, which limits electron acceleration. Therefore, in our study, the STOV field with $l = 1$ proves to be a more suitable structure for generating isolated attosecond $\gamma$-ray pulse.

In conclusion, this study proposed an all-optical scheme for generating isolated attosecond $\gamma$-ray pulse with TOAM using an intense CP STOV laser on a wire target, as demonstrated through 3D PIC simulations. We found that an IAES, characterized by $\sim$ 300 as duration, $\sim$ 0.2 nC charge, $\sim$ 2$\degree$ divergence angle, and $\sim$ 600 MeV cutoff energy, is initially modulated by the spatiotemporal singularity fields at the center of the STOV laser. This pulse subsequently collides with the reflected Gaussian-like front of the planar target, leading to the NCS process that produces an isolated attosecond $\gamma$-ray pulse with a duration of $\sim$ 300 as, a divergence angle of $\sim$ 4$\degree$, and an extraordinary brilliance of $\sim 5\times 10^{24}$ photons/s/mm$^2$/mrad$^2$/0.1\%BW at 1 MeV) $\gamma$-ray pulse. Unlike previous $\gamma$-ray sources driven by traditional Gaussian lasers, this novel $\gamma$-ray pulse carries a distinct TOAM, introducing revolutionary capabilities for relativistic modulation that could enable ultra-high time-resolution observations, selective nuclear excitation, and enhanced detection in nuclear science.

\emph{Acknowledgments}: We thank Prof. B.-S. Xie, Dr. C.-W. Zhang and Dr. H. Zhang for fruitful discussions. This study is supported by the Natural Science Foundation of China (Grant No. 12075306), Natural Science Foundation of Shanghai (Grant No. 22ZR1470900), Key Research Programs in Frontier Science (Grant No. ZDBSLY-SLH006), Strategic Priority Research Program of the Chinese Academy of Sciences (Grant No. XDA0380000), Shanghai Science and Technology Committee Program (Grant No. 22DZ1100300).

\end{document}